\documentclass[12pt,twoside]{article}
\input{epsf.sty}

\def\mypagenumber{1}
\def\mydate{September 23, 1998}
\def\myend{\end{document}}

\def\Journal#1#2#3#4{{#1}{\bf #2} (#3) #4}
\def\NPB{{\em Nucl.\ Phys.} B}
\def\PLB{{\em Phys.\ Lett.} B}
\def\PRL{\em Phys.\ Rev.\ Lett. }

\def\PRD{{\em Phys.\ Rev.} D}
\def\AP{{\em Ann.\ Phys.\ (N.Y.)} }

\def\cmp{{\em Com.\ Math.\ Phys.}}

\normalsize
\let\oldtheequation=\theequation
\def\doteqs#1{\setcounter{equation}{0}
            \def\theequation{{#1}.\oldtheequation}}
\newcounter{sxn}
\def\sx#1{\addtocounter{sxn}{1} \vskip 1.cm  \goodbreak
\noindent{\large\bf\leftline{\thesxn.~~#1}} \nobreak \vskip -.5cm}
\def\sxn#1{\sx{#1} \doteqs{\thesxn}}

\newcounter{axn}

\date{}

\newdimen\mybaselineskip
\newcommand{\beeq}{\begin{equation}}
\newcommand{\eneq}{\end{equation}}
\newcommand{\be}{\begin{eqnarray}}
\newcommand{\ee}{\end{eqnarray}}
\newcommand{\bpic}{\begin{picture}}
\newcommand{\epic}{\end{picture}}

\def\dd{\partial}
\def\la{\raise.16ex\hbox{$\langle$} \, }
\def\ra{\, \raise.16ex\hbox{$\rangle$} }
\def\go{\rightarrow}

\def\next{{~~~,~~~}}
\def\ie{{\it i.e., \ }}

\def\psibar{ \psi \kern-.65em\raise.6em\hbox{$-$} }
\def\mbar{ m \kern-.78em\raise.4em\hbox{$-$}\lower.4em\hbox{} }

\def\phibar{{\bar\phi}}
\def\fbar{{\bar f}}

\def\Abar{{\overline A}}

\def\ep{\epsilon}

\def\vx{{\vec x}}

\def\n@space{\nulldelimiterspace=0pt \mathsurround=0pt }
\def\huge#1{{\hbox{$\left#1\vbox to 20.5pt{}\right.\n@space$}}}

\def\myskip{\noalign{\kern 8pt}}
\def\myeqspace{\noalign{\kern 10pt}}

\def\boxit#1{$\vcenter{\hrule\hbox{\vrule\kern3pt
    \vbox{\kern3pt\hbox{#1}\kern3pt}\kern3pt\vrule}\hrule}$}
\def\bigbox#1{$\vcenter{\hrule\hbox{\vrule\kern5pt
     \vbox{\kern5pt\hbox{#1}\kern5pt}\kern5pt\vrule}\hrule}$}

\def\ignore#1{{}}

\begin{document}

\bibliographystyle{unsrt}
\footskip 1.0cm

\thispagestyle{empty}
\setcounter{page}{\mypagenumber}

%{\baselineskip=10pt \parindent=0pt \small
%\mydate 
%}                
             
\begin{flushright}{\mydate ~ (corrected.)
\\  UMN-TH-1626/98  , TPI-MINN-98/11-T \\
hep-th/9808045}
\\
\end{flushright}

\vspace{2.5cm}
\begin{center}
{\LARGE \bf {Complex  Monopoles}}\\
\vskip .8cm 
{\LARGE \bf {in the Georgi-Glashow-Chern-Simons Model}}\\
\vspace{3cm}
{\large Bayram Tekin, Kamran Saririan and Yutaka Hosotani}\\

\vspace{.5cm}
{\it School of Physics and Astronomy, University of Minnesota}\\ 
{\it  Minneapolis, MN 55455, U.S.A.}\\ 
\end{center}

\vspace*{2.5cm}
%\baselinestretch{2.0}

%\normalsize

\begin{abstract}
\baselineskip=18pt
We investigate  the  three dimensional 
Georgi-Glashow model with a Chern-Simons term.  
We find that there exist complex monopole
solutions of finite action.  They dominate the path integral and 
disorder the Higgs vacuum, but electric charges are not confined.
Subtleties in the gauge fixing procedure in the path integral and 
issues related to Gribov copies are noted.
\end{abstract}

\vfill

PACS: ~ 11.10.Kk; 11.15.Kc; 11.15.Tk

Keywords: ~  Monopoles, Chern-Simons theory, Confinement

%\end{titlepage}
 
\newpage

%\setcounter{page}{1}

%\textheight=20cm
%\headsep=0.75cm
%\vsize=20cm

%%%%%%%%%%%%%%%%%%%%%%%%%%%%%%%%%%%%%%%%%%%%%%%%%%%%%%%%%%%%%%%%%
\normalsize
\baselineskip=22pt plus 1pt minus 1pt
\parindent=25pt
%\vspace*{5mm}

\sxn{Introduction}

Years ago Polyakov  showed 
that in the three-dimensional Georgi-Glashow  model,
or more generally in the compact QED$_3$,  monopole 
configurations in the Euclidean space make dominant contributions 
in the functional integral for the  confinement of
electric charges \cite{Polyakov}. The logarithmic potential between two
electric probe charges is converted to a linear potential 
in the background of a monopole gas, leading to the linear confinement.
The Higgs vacuum is disordered.  The expectation value of the 
triplet Higgs field vanishes ($\la \vec \phi \ra =0$), whereas all
components of the
$SU(2)$ gauge fields acquire masses.  The long range order in 
the Higgs vacuum is destroyed by monopole configurations.

Gauge theory in three dimensions can
accommodate a purely topological term, the Chern-Simons term,  which
affects the equations of motion. It gives a topological mass to   
gauge fields \cite{Deser}.  In the Georgi-Glashow model, even in perturbation 
theory,
the unbroken $U(1)$ gauge field also becomes massive, and thus the issue of
the linear confinement disappears in the presence of the Chern-Simons
term;  there is no long-range force in the Georgi-Glashow-Chern-Simons 
(GGCS) model to start with.  The electric
flux is not conserved.  It does not matter for the issue of the 
confinement whether or not monopole configurations dominate in the 
functional integral.

 Once the Chern-Simons term is added the action becomes complex in
the Euclidean space. Many authors have shown that there is no real
monopole-type field
configuration of a finite action which solves the Euler equations 
\cite{Pisarski}-\cite{Diamantini}.
This fact
has been interpreted as indicating the irrelevance of (real) monopole
configurations in the model.

There remain a few  puzzles. Although there is no  confinement in the presence
of the Chern-Simons term,  there remains the issue of the long-range
order in the Higgs vacuum.  How can the Higgs vacuum be disordered
if  monopole configurations are totally irrelevant?  Does the 
expectation value $\la \vec \phi \ra$ become nonvanishing and
the long-range order is restored once the  Chern-Simons term is added?
Also, if the theory allows complex monopole configurations, does their 
contribution to the partition function vanish as in the real case? 
There are also subtle questions  related to the gauge invariance as 
well as Gribov copies in various gauges in the GGCS model which, to 
our knowledge, have not yet been answered.

We shall re-examine these considerations in the context of 
complex monopole solutions to show that the Higgs vacuum 
remains disordered.  Although
there are no real monopole configurations which solve the Euler
equations, there exist complex monopole configurations which 
extremize the Euclidean action in a fixed gauge.  
They dominate the
functional integral in quantum theory and destroy the long-range order in
the Higgs vacuum. 
The effect of Gribov copies is also re-examined.  What we mean by complex
monopoles will be clear in the text but for now we should state that these
are complex-valued solutions to the equations of motion. The
non-abelian field strengths are complex but the $U(1)$ 't Hooft
field strengths are real and exactly those of 
 a magnetic monopole. With the $U(1)$ projection our 
complex monopoles can be interpreted as the topological excitations 
characterized  by the the group $\Pi_2(SO(3)/U(1))= Z$.

It is worthwhile to recall the correspondence between compact QED$_3$ and 
the Josephson junction system in the 
superconductivity \cite{Hosotani}.  The normal barrier region 
sandwiched by two bulk superconductors becomes superconducting
due to supercurrents flowing through the barrier.  The
three-dimensional compact QED is related to the Josephson 
junction system by the electro-magnetic dual transformation.
The $U(1)$ field strengths $(E_1, E_2, B)$ in the Georgi-Glashow model
correspond to $(B_1, B_2, E_3)$ in the Josephson junction.
Electric charges in the Georgi-Glashow model are magnetic charges
inserted in the barrier in the Josephson junction.  If there were no
supercurrents, the magnetic flux  between a pair of magnetic
monopole and anti-monopole inserted in the barrier forms dipole
fields, giving a logarithmic potential between the pair.  However,
due to supercurrents the magnetic flux is squeezed to form a
Nielsen-Olesen vortex giving rise to a linear potential.

Monopoles (instantons) in compact  QED$_3$ are supercurrents
in the Josephson junction.  Polyakov introduced a collective field
$\chi$ which mediates interactions among monopoles.  The field $\chi$
corresponds to the difference  between the phases of the
Ginzburg-Landau order parameters $\Psi_{\rm GL}$
in the bulk superconductors on both 
sides of the barrier in the Josephson junction.  Both $\chi$ and
$\delta ({\rm arg} \Psi_{\rm GL})$ satisfy the same sine-Gordon type
equation.

Now we add the Chern-Simons term in the Georgi-Glashow model.  
At the moment we haven't understood what kind of an additional
interaction in the Josephson junction system corresponds 
to the Chern-Simons term in the Georgi-Glashow model.
It could be a $\theta F_{\mu\nu} \tilde F^{\mu\nu}$ term in the
superconductors on both sides.  Normally a 
$\theta F_{\mu\nu} \tilde F^{\mu\nu}$ term is irrelevant in QED.
However, if the values of $\theta$ on the
left and right sides are different, this term may result in a physical
consequence, which may mimic the effect of the Chern-Simons term in
the Georgi-Glashow model.    

If monopoles are irrelevant in the presence of the Chern-Simons term,
it would imply that suppercurrents cease to flow across the barrier
in the corresponding Josephson junction.  Although we have not found
the precise analogue in the Josephson system yet, and therefore we cannot
say anything definite by analogy, we feel that it is very puzzling 
if suppercurrents suddenly stop to flow.  Monopoles should remain
important even in the presence of the Chern-Simons term.  
The correspondence between the  Chern-Simons theory and the 
Josephson junction arrays has been  discussed in ref.\
\cite{Diamantini2}.

As we shall discuss below, there is a subtle issue in quantizing a
Chern-Simons theory.   The arguments below are based on the Chern-Simons
theory in a fixed gauge.  In the path integral a gauge condition
restricts functional space to be integrated.  Within this subspace
complex monopole solutions are found.  This is a delicate issue as
the Chern-Simons term is not gauge invariant.

\sxn{The model}
The action for the scalar fields interacting with  the gauge fields in the
three-dimensional Euclidean space is given by 
$S= S_0 + S_{\mbox{cs}} + S_h$
where 
\be
S_0 &=& -{1\over 2g^2}\int d^3x \mbox{tr}\left(F_{\mu \nu}F^{\mu 
\nu}\right)   \cr
\noalign{\kern 10pt}
S_{\mbox{cs}} &=& -{i\kappa\over g^2}\int d^3x \epsilon^{\mu \nu
\lambda}\mbox{tr}\left(A_\mu \dd_\nu A_\lambda +{2\over 3} A_\mu
A_\nu A_\lambda \right)  \cr
\noalign{\kern 10pt}
S_h &=& {1\over g^2}\int d^3x \left[{1\over 2}(D_\mu h^a)^2 +{\lambda
\over  4}(h^a h^a - v^2)^2 \right] 
\label{model1}
\ee
We adopt the notation   $A_\mu = {i\over 2} A^a_\mu \tau^a$, 
$F_{\mu\nu} = \dd_\mu A_\nu  -\dd_\nu A_\mu +[A_\mu, A_\nu]$,
$h = {i\over 2} h^a \tau^a$, and $D_\mu h = \dd_\mu h + [A_\mu, h]$. 
The classical equations of motion are
\be
&D_\mu F^{\mu\nu} +{i\over 2}\kappa \epsilon^{\nu\lambda\mu}F_{\lambda
\mu} =  [h,D^\nu h] ~~,& \cr
&D_\mu D^\mu h = -\lambda(\mbox{tr}h^2 + m^2) h ~~.&
\label{EulerEq1}
\ee

Notice that $S_{\mbox{cs}}$ is pure imaginary in the Euclidean space
for real gauge field configurations. Equations of motion become complex. 
To define a quantum theory,
one needs to fix a gauge. 
We would like to find   dominant field configurations.   Polyakov, in the
theory where $\kappa=0$, showed that monopole  configurations are 
essential in removing the degeneracy of the vacuum and lead to the
linear confinement of electric charges. For $\kappa \neq 0 $ a real
monopole configuration is
shown to have an infinite action which makes its contribution vanish
 in the vacuum.   Pisarski \cite{Pisarski} interpreted this as the
confinement of monopole.  Fradkin and Schaposnik  \cite{Fradkin} 
consider a theory where Chern-Simons term is induced as a one loop
effect after integrating   fermions.  In this case  
the t'Hooft-Polyakov real monopole \cite{Polyakov,tHooft} minimizes the
classical equations of motion but  not at the one loop level. Fradkin
and Schaposnik deform the real t'Hooft-Polyakov monopole in the complex
configuration space.  They have concluded that a monopole-antimonopole
pair is bound together by a linearly growing potential. 

It is clear that more careful analysis is necessary to find the absence or
existence of complex monopole configurations and their implications.
In the context of path integrals, stationary points of the exponent in
the integrand may be located generically at complex points, though
the original integration over field configurations is defined along 
the real axis.  In the saddle-point method for integration such
saddle-points give dominant contributions in the integral.  Complex
monopoles can be vital in disordering the vacuum.

When $v\not=0$ in  (\ref{model1}), 
the perturbative vacuum manifold is $SO(3)/U(1)$ .   
Following  t'Hooft and Polyakov, we make the 
spherically symmetric monopole ansatz which breaks 
$SO(3)_{\rm gauge} \times SO(3)_{\rm rotation}$ to $SO(3)$. 
\be
&&h^a(\vec{x})=\hat x^a h(r) \cr
&&A^a_\mu(\vec{x})= {1\over r} \left[ \epsilon_{a\mu
\nu}\hat{x}^\nu(1-\phi_1)+ \delta^{a\mu}\phi_2 +(r A-\phi_2)\hat{x}^a 
\hat{x}_\mu \right]       
\label{configuration1}       
\ee 
where $\hat x^a = x^a/r$.
Field strengths are
\be
F_{\mu\nu}^a &=&
 {1\over r^2}
  \ep_{\mu\nu b}  \hat x^a \hat x^b (\phi_1^2 + \phi_2^2 -1)
+ {1\over r} (\ep_{a\mu\nu} -  \ep_{\mu\nu b}  \hat x^a \hat x^b )
(\phi_1' + A\phi_2) \cr
\noalign{\kern 10pt}
&& \hskip 4cm + (\delta^{a\nu} \hat x^\mu-\delta^{a\mu} \hat x^\nu) 
{1\over r}(\phi_2' - A\phi_1)
\label{non-abelian field strength}
\ee

Classical equations of motion are
\beeq
{1\over r^2} {d\over dr} (r^2 h') - \lambda (h^2 - v^2) h -
{2\over r^2} (\phi_1^2 + \phi_2^2) h = 0 
\label{EqMotion1}
\eneq
for the Higgs field and 
\begin{eqnarray}
&&\phi_1 \phi_2' - \phi_2 \phi_1' -A(\phi_1^2 +\phi_2^2) + 
{i \over 2}\kappa (1-\phi_1^2 -\phi_2^2) = 0  \label{EqMotion2} \\
&& (\phi_1' +A \phi_2)' +A(\phi_2'-A\phi_1) + {\phi_1 \over r^2}
(1-\phi_1^2 -\phi_2^2) + i\kappa (\phi_2'- A\phi_1) -h^2\phi_1 = 0 
 \label{EqMotion3}\\
&& (\phi_2' -A \phi_1)'- A(\phi_1'+A\phi_2 ) + {\phi_2 \over r^2}
 (1-\phi_1^2 -\phi_2^2) - i\kappa (\phi_1'+ A\phi_2)  
-h^2\phi_2 = 0  \label{EqMotion4}
\end{eqnarray}
for the gauge fields.
Finite action solutions to these equations are necessarily complex for
$\kappa
\not= 0$.  Note that it follows from Eqs.\ (\ref{EqMotion3}) and
(\ref{EqMotion4}) that
\beeq
\phi_1 \phi_2' - \phi_2 \phi_1' -A(\phi_1^2 +\phi_2^2) + 
{i \over 2}\kappa (1-\phi_1^2 -\phi_2^2) = {\rm const}
  \label{EqMotion5}
\eneq
Eq.\ (\ref{EqMotion2}) gives an additional information that
the constant in Eq.\ (\ref{EqMotion5}) is 0.

However, due caution is necessary in writing down equations of motion 
in quantum theory.  The Chern-Simons term is not gauge invariant.
Even if the action is varied in a pure gauge direction, it may
change.  In other words the action need not be stationary under such
variations.
In the monopole ansatz one combination of Eqs.\
(\ref{EqMotion2}) - (\ref{EqMotion4}) corresponds to such variations.
If one fixes the gauge first, this particular equation does not
follow from the gauge-fixed action.

This  becomes clearer when equations are derived from the action
written in terms of the monopole ansatz.
The action is  then written as follows (using eq. (\ref{configuration1})):
\be
S &=& {4\pi\over g^2} \int_0^\infty dr \, \Bigg\{(\phi_1' + A\phi_2)^2 +
(\phi_2' - A\phi_1)^2 + {1\over 2r^2} (\phi_1^2 + \phi_2^2 -1)^2 \cr
&&\hskip 2cm
+i \kappa \Big[ \phi_1' \phi_2 - \phi_2'(\phi_1 -1) + A(\phi_1^2 +
\phi_2^2 -1) \Big] \cr
&&\hskip 2cm 
+{r^2\over 2} h'^2 + h^2 (\phi_1^2+\phi_2^2) + {\lambda r^2\over 4}
(h^2-v^2)^2 \Bigg\} 
\label{action2}
\ee
The regularity at the origin and the finiteness of the action place
boundary conditions
\be
{\rm at} ~ r=0 : \hskip .2cm 
   &&h=0 \next  \phi_1 = 1 \next  \phi_2 =0 \cr
{\rm at} ~ r=\infty : && h=v \next \phi_1 =0 \next \phi_2 = 0 \next A=0 ~~.
\label{BC1}
\ee
In the original 't Hooft-Polyakov \cite{Polyakov,tHooft} monopole
solution, $\phi_2(r)=A(r)=0$. 

For  $v \neq 0 $ the unbroken $U(1)$ is parametrized by  residual
gauge transformations generated by $\Omega = \exp \Big\{ 
{i\over 2} f(r)\, \hat x^a\sigma^a \Big\}$  where $f(0)= 0$. 
Under this gauge transformation the fields transform as
\be
\phi_1 &\go& + \phi_1 \cos f +\phi_2\sin f \cr
\phi_2 &\go&  -\phi_1 \sin f +\phi_2\cos f \cr
A &\go& A - f' 
\label{transform1}
\ee

The equations of motion are invariant under this transformation but the
action (\ref{action2}) is not. 
Under a more general gauge transformation $A \go \Omega A \Omega^{-1}
+ g^{-1} \Omega d \Omega^{-1}$, 
\beeq
S_{CS} \go S_{CS} + {i\kappa\over g^2} \int 
   {\rm tr~} d(A \wedge d\Omega^{-1} \Omega)
+ {i\kappa\over 3g^2} \int {\rm tr~} d\Omega^{-1} \Omega
   \wedge d\Omega^{-1} \Omega \wedge d\Omega^{-1} \Omega ~~.
\label{GT1}
\eneq
If the theory is defined on $S^3$, the first term vanishes.
The last term is the winding number of the mapping $\Omega$.
This leads to the celebrated quantization condition
$4\pi\kappa/g^2 = n =$ an integer \cite{Deser}.

On $R^3$, however, the first term does not vanish for monopole
configurations.  Under (\ref{transform1}) the second and last terms
in (\ref{GT1}) are $(4\pi i \kappa/g^2) \sin f(\infty)$ and
$(4\pi i \kappa/g^2) ( f(\infty) -\sin f(\infty))$, respectively.
Hence
\beeq 
S \go S + {4\pi i\kappa\over g^2}  f(\infty)  ~~~.
\label{transform2}
\eneq

If the action is viewed as a functional of all $h$, $A$, $\phi_1$, 
and $\phi_2$ and varied with respect to those fields, then one would 
obtain Eqs.\ (\ref{EqMotion1}) - (\ref{EqMotion4}).  It is obvious,
however, that Eq.\ (\ref{EqMotion2}) and the boundary condition
(\ref{BC1}) are incompatible at $r=\infty$. There would be no solution
of a finite action.   Instead, one
might fix a gauge first and vary the action within the gauge chosen.
Indeed, this is what is done in quantum
theory either in the canonical formalism or in the path integral
formalism.  In the path integral we start with
\beeq
Z = \int {\cal D} A {\cal D} h \, \Delta_F(A) \delta[F(A)] \,
  e^{- S} ~~.
\label{PIformula1}
\eneq
What we are looking for is a field configuration which extremizes
$S$ within the gauge chosen;
$\delta S \big|_{F(A)=0}$.
If the action is manifestly gauge invariant, the order of two operations,
fixing a gauge and extremizing $I$, does not matter.  However, in the 
presence of the Chern-Simons term, the action is not manifestly gauge
invariant.  As we shall see explicitly in the following section, 
two configurations $A^{(j)}$ ($j=1, 2)$ determined by $\delta S
\big|_{F_j(A)=0}$ are physically different in general, i.e.\ $A^{(1)}$ is
not on a gauge orbit of $A^{(2)}$.  

Before closing this section, we would like to remark that there are 
three  approaches in defining the path integral.  The issue
is how to define the configuration space $A_\mu$ in (\ref{PIformula1}).
The first possibility is to restrict gauge orbits such that transformation
function $\Omega$  be $S^3$-compatible, namely
$\Omega|_{r=\infty}=\Omega_\infty$.  For (\ref{transform1}) it implies 
that $f(\infty)=2\pi p$ where $p$ is an integer.  With this $e^{-S}$
becomes gauge invariant.  However, this restriction leads to
 conflict with gauge fixing.  Suppose that a gauge configuration 
is given.  Now one fixes a gauge.  However, a gauge orbit of the given
configuration may not intersect the gauge fixing condition.  For
instance,  if $\int_0^\infty dr \, A \not= 2\pi p$, then the 
configuration cannot be represented in the radial gauge.  If 
$\phi_2/\phi_1 |_{r=\infty} \not= 0$, then the configuration 
cannot be represented in the unitary gauge.  If there is no $f(r)$
satisfying
\beeq
f'' + {2\over r} f' - {2\over r^2}
\big\{ \phi_1 \sin f + \phi_2 (1 - \cos f) \big\} 
= A' + {2A\over r}  - {2 \phi_2\over r^2}~~~,
\eneq
then the configuration cannot be represented in the radiation gauge.
Put it differently,  the gauge fixing procedure removes a part of 
physical gauge configurations. This approach is not acceptable.

The second possibility is to impose no restriction on $\Omega$.
In the case $e^{-S}$ is not gauge invariant in general.   In this
paper we adopt this approach to find consequences.

The third possibility is to restrict gauge field configurations
$A_\mu$  to be $S^3$ compatible.  This excludes monopole ansatz
(\ref{configuration1}) entirely.  With this restriction the total
monopole charge must vanish.  This is certainly a legitimate
approach, and there occurs no problem of the gauge invariance of the 
theory.  One has to address a question why the spacetime needs to be
compactified from $R^3$ to $S^3$ in defining a theory.    We leave this
possibility for future consideration.

\sxn{Gauge choices}

There are four gauge choices  which are typically considered:

\bigskip

\noindent {\bf (i) Radial gauge} 
  ($A = \hat x^a \hat x^\mu A^a_\mu =0$).

As is obvious from (\ref{transform1}), this gauge choice is always
possible.  $h$, $\phi_1$ and $\phi_2$ are independent fields.
Equations  derived by extremizing the action (\ref{action2}) are
Eq.\ (\ref{EqMotion1}) and
\be
&&\phi_1'' + {1\over r^2} (1-\phi_1^2  -\phi_2^2) \phi_1
 + i\kappa \phi_2' - h^2 \phi_1 = 0 \label{EqMotion6} \\
&&\phi_2'' + {1\over r^2} (1-\phi_1^2  -\phi_2^2) \phi_2
 - i\kappa \phi_1' - h^2 \phi_2 = 0 
\label{EqMotion7}
\ee
Eq.\ (\ref{EqMotion6}) and (\ref{EqMotion7}) are obtained by naively
setting $A=0$ in (\ref{EqMotion3}) and (\ref{EqMotion4}).  Eqs.\
(\ref{EqMotion6}) and (\ref{EqMotion7}) imply (\ref{EqMotion5})
but not necessarily (\ref{EqMotion2}). The left hand side of
(\ref{EqMotion2}) does not vanish in general. 

\bigskip 
\noindent {\bf (ii) Unitary gauge ($\phi_2 =0$).}

This gauge was employed in ref. \cite{Pisarski}.  Equations which follow
are  (\ref{EqMotion1}) and
\be 
&&\phi_1^2 A - {i\over 2} \kappa (1 - \phi_1^2) =0
   \label{EqMotion8} \\
&&\phi_1'' - A^2 \phi_1 + {1\over r^2} (1-\phi_1^2) \phi_1 - i\kappa
A\phi_1 =0
  \label{EqMotion9}
\ee
Eq.\ (\ref{EqMotion8})  follows from (\ref{EqMotion4}) by setting
$\phi_2= 0$.

Eq.\ (\ref{EqMotion8}) is incompatible with the boundary condition (\ref{BC1})
for $\kappa\not= 0$.  Even for configurations which do not
satisfy the Euler equations,  the unitary gauge may not be possible.  Suppose
that a configuration $(h, \phi_1, \phi_2, A)$ satisfies the boundary
condition (\ref{BC1}), yielding a finite action.  
To bring it to the unitary gauge one has to choose
$\tan f = (\phi_2/ \phi_1)$ in (\ref{transform1}).
The action changes by a finite amount
\beeq
 {4\pi\kappa\over g^2} ~
\tan^{-1}{\phi_2\over \phi_1}\Bigg|^{r=\infty}_{r=0} ~~.
\label{transform3}
\eneq
The new $A$ satisfies the boundary condition
only if $f'=(\phi_1 \phi_2' - \phi_1' \phi_2)/(\phi_1^2 + \phi_2^2)$
vanishes at $r=\infty$, which is not generally satisfied.

\bigskip

\noindent {\bf (iii) Radiation gauge ($\dd_\mu A^a_\mu=0$).}

The gauge condition implies that
\beeq
A' + {2\over r} \Big( A - {\phi_2\over r} \Big) = 0 \next
A(r) = {2\over r^2} \int_0^r dr \, \phi_2(r) 
\label{radiation1}
\eneq
The action is viewed as a functional of $h$, $\phi_1$ and $\phi_2$.
Equations derived are Eq.\  (\ref{EqMotion1}), (\ref{EqMotion3}) and
\begin{eqnarray}
&& (\phi_2' -A \phi_1)'- A(\phi_1'+A\phi_2 ) + {\phi_2 \over r^2}
 (1-\phi_1^2 -\phi_2^2) - i\kappa (\phi_1'+ A\phi_2)  
-h^2\phi_2 \cr
&&\hskip 1cm - 2 \int_r^\infty du \,
{1\over u^2} \Big\{ \phi_2 \phi_1' - \phi_1 \phi_2' + A(\phi_1^2 +
\phi_2^2) 
+ {i\over 2} \kappa (\phi_1^2 +\phi_2^2 - 1) \Big\} =0
   \label{EqMotion10}
\end{eqnarray}
Here $A$ is expressed in terms of $\phi_2$ by (\ref{radiation1}).

There is residual gauge invariance.  The gauge condition is respected 
by the transformation (\ref{transform1}), provided 
$f(r)$  obeys the Gribov equation \cite{Gribov}: 
\beeq
f'' + {2\over r} f' - {2\over r^2} \Big\{ \phi_1 \sin f
+ \phi_2 (1-\cos f) \Big\} = 0 
\label{radiation2}
\eneq
or
\beeq
\ddot{f}  + \dot{f} - 2 \Big\{ \phi_1 \sin f
+ \phi_2 (1-\cos f) \Big\} = 0 ~~.
\label{radiation3}
\eneq
Here a dot represents a  derivative with respect to $t=\ln r$.

As was pointed out by Gribov, Eq.\ (\ref{radiation3}) is an equation for
a point particle in a potential $V=2\big\{ \phi_1 \cos f - \phi_2(f - \sin
f)\big\}$ with friction force.  For $\phi_1=1$ and $\phi_2=0$, which
corresponds to the trivial vacuum configuration $A^a_\mu=0$ in 
(\ref{configuration1}), there appear three types of solutions.  
With the initial condition $f|_{r=0}=0$, (i) $f(r)=0$, (ii)
$f(\infty)=\pi$, or (iii) $f(\infty) = -\pi$.  The last two are the
Gribov copies  (see Fig.\ 1).

\begin{figure}[ht]\centering
 \leavevmode 
\mbox{
\epsfysize=3in \epsfbox{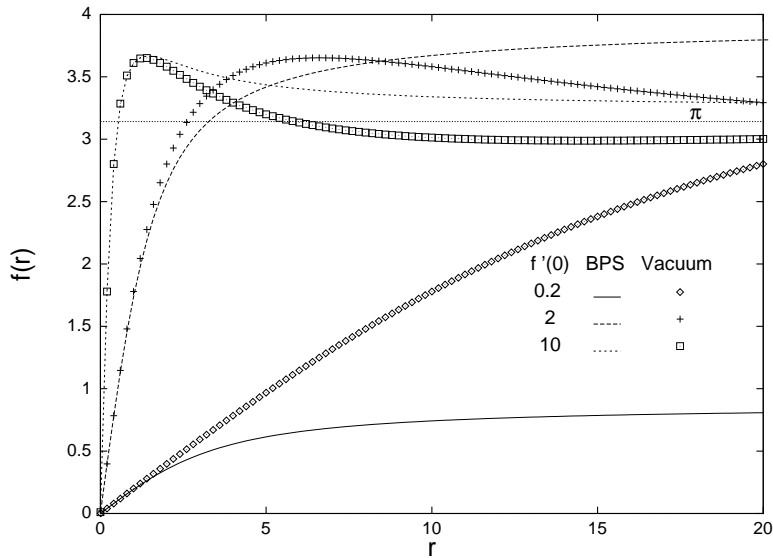}}
\caption{Solutions $f(r)$ to the Gribov Eq.\ (3.8). 
Solid lines correspond to the BPS monopoles with various values of $f'(0)$
whereas points correspond to the vacuum. For the vacuum $f(\infty) =\pi$
for positive $f'(0)$.} 
\end{figure}

It is interesting to consider the relevance of the 
't Hooft-Polyakov monopole configurations $\phi_2=0$ 
but $\phi_1\not= 0$ ($\phi_1(0)=1$ and $\phi_1 \go 0$ as 
$r\go \infty$).\footnote{
 Although this is clearly not a solution to the theory including
the CS term, this exercise illustrates the possibility that summing of 
the Gribov copies may not lead to the cancellation of the monopole-type 
contribution.}
  In terms of the Gribov equation, this case corresponds to a 
particle moving 
in a time-dependent potential.  As the potential  becomes exponentially
small for large $t$, there can be a continuous family of solutions
parameterized by the value of $f(\infty)$.   The asymptotic value
$f(\infty)$ depends of the initial slope $f'(0)$.  In the BPS limit it 
ranges from $0$ to $3.98$.  See Fig.\ 2.  For $|f'(0)| \ll 1$,
$|f(r)|$  remains
small.  For $|f(r)| \gg 1$, $f(r)$ approaches an asymptotic value
before $\phi_1$ and $\phi_2$ make sizable changes, {\it i.e., } 
$f(r)$ behaves as in the vacuum case.  The maximum value for
$|f(\infty)|$ is attained for $f'(0)=\pm 2.62$.

\begin{figure}[ht]\centering
 \leavevmode 
\mbox{
\epsfysize=2.3in \epsfbox{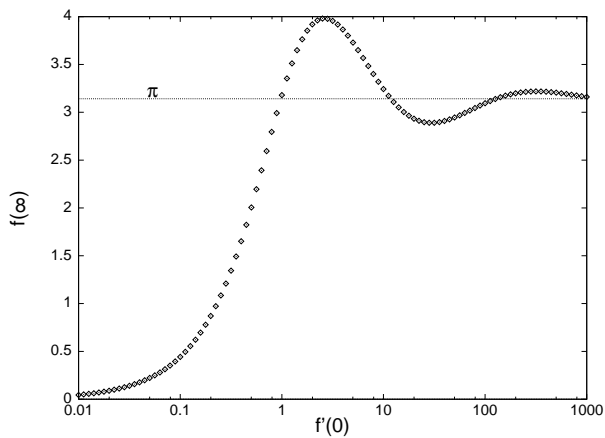}}
\caption{$f(\infty)$ for BPS monopole vs. initial slope ($f'(0)$).}
\end{figure}

These Gribov copies in the radiation gauge lead to an important 
consequence in the Chern-Simons
theory.  Under the gauge transformation  the action changes as
\be
S \go  S +i{4\pi \kappa \over g^2} ( f(\infty) -f(0) ) ~~.
\ee
Recall that the Chern-Simons coefficient is quantized, i.e.\ 
$4\pi \kappa/g^2$ is an integer.
If $f(\infty)$ takes an arbitrary value, or takes  all values in the
interval
$[-\pi,+\pi]$, then the integration over the parameter $f(\infty)$ in the
path integral could kill contribution coming from the monopole
configuration as argued by Affleck {\it et. al.} 
\cite{Affleck}.\footnote{In
ref. \cite{Affleck}, the parameter  $f(\infty)$ is interpreted 
as the collective 
coordinate of the monopole configuration, and is therefore integrated
over.  Here, we alternatively 
associate  $f(\infty)$ with the Gribov copies within a given gauge, 
corresponding to 
the solutions of eq. (\ref{radiation2}). In this interpretation, one
sums over the 
Gribov copies, \ie one integrates over the parameter  $f(\infty)$. }
  A close examination of 
Eq.\ (\ref{radiation2})  reveals that this is not the case.  As
mentioned above, $f(\infty)$ for the 't Hooft-Polyakov monopole 
ranges from $-3.98$ to $3.98$, depending on
$f'(0)$.   There seems no reason for expecting the cancellation.

\bigskip

\noindent {\bf (iv) Temporal gauge ($A^a_3 =0$)}

This gauge destroys spherical symmetry in the Euclidean space, but
allows physical interpretation in the Minkowski spacetime.

\bigskip

We are going to show that there are complex monopole solutions in
certain gauges.  We argue also that in the case of the radiation gauge 
Gribov copies do not lead to cancellation.   When $\phi_2$ is
complex in (\ref{radiation2}), a solution $f(r)$ to Eq.\
(\ref{radiation2}) necessarily becomes complex.  In other words the
transformation specified with
$f(r)$ is not a gauge transformation.  However, one can show that
there are solutions $f(r)$ to Eq.\ (\ref{radiation2}) in which $f(0)=0$
and  $f(\infty)$ is real.  

In terms of a gauge  invariant quantity
\be
\eta ~ &=& (\phi_1 + i \phi_2)e^{- i\int^r A(r) dr}  \cr
\eta^* &=& (\phi_1 - i \phi_2)e^{+ i\int^r A(r) dr}  
\ee
the action (\ref{action2}) simplifies to
\be
S&=& {4\pi\over g^2} \int_{0}^{\infty} dr \Big\{ {\eta'^*}{\eta'}
    +{1\over 2 r^2}(1- \eta\eta^*)^2 - {\kappa\over 2} (\eta'\eta^* 
    -\eta {\eta'^*}) +i \kappa A \nonumber \\
\noalign{\kern 10pt} 
&& \hskip 2cm +{r^2\over 2} (h')^2 + h^2 \eta \eta^* + 
{\lambda r^2\over 4}(h^2-v^2)^2  \Big\} ~.
\ee
Note that $\int_0^\infty dr \, A(r)$ implicitly depends on $\eta$ 
and $\eta^*$.

\sxn{Complex monopole solutions}

As discussed in the previous section, the gauge is fixed in quantum
theory.  The aim of this section is to show the existence of complex
monopole solutions in certain gauges.  We shall see a subtle relation
among different gauges.   A solution in one gauge can be
transformed into another gauge by a ``complex'' gauge transformation, 
but the transformed configuration may not be a solution in the new
gauge.  This is a reflection of the gauge non-invariance of the original theory
in the presence of the Chern-Simons term.
It implies that relevant field configurations in the path integral
may appear different, depending on the gauge.  This raises a question
if all gauges are equivalent as they should.
A related issue has been analyzed in the temporal gauge in ref. \cite{Dunne}.
It has been shown that  implementation of constraints before
quantization  does not yield the correct physics in Chern-Simons theory. 

We note that the  t'Hooft   $U(1)$ field strength is defined by
\be
F_{\mu\nu} = {h^a\over h} F_{\mu\nu}^a - {1\over h^3} \epsilon_{abc}
             h^a(D_\mu h)^b (D_\nu h)^a
\ee
With our ansatz it becomes
\be
F_{\mu\nu}= - {\ep_{\mu\nu a} \hat x^a \over r^2}~~~. 
\label{Abelian}
\ee
It is independent of the details of $(\phi_1, \phi_2, A)$.
In the solution for $\kappa \not= 0$, $\phi_2$  and therefore
non-abelian field strengths generally become complex.
However, the 't Hooft tensor (\ref{Abelian}) remains real with
the standard  Abelian field strength representing a magnetic
monopole.  We call the solution a complex monopole.
We also recall that in the BPS limit $\lambda=\kappa=0$
the solution is given by 
\beeq
h(r)= v \coth(v r) -{1\over r}  \next
\phi_1(r) = {v r\over \sinh(v r) } \next 
\phi_2=A=0 ~~~.
\label{BPS1}
\eneq

\bigskip

\noindent {\bf (i) Radial gauge ($A=0$)}

In this gauge there is no Gribov copy.  Equations to be solved are
Eqs.\  (\ref{EqMotion1}), (\ref{EqMotion6}), and (\ref{EqMotion7}). 
There is a solution in which $h$ and $\phi_1$ are real, but $\phi_2$
is pure imaginary.  An appropriate ansatz is
\beeq
\phi_1(r) = \zeta(r) \cosh {\kappa r\over 2} \next
\phi_2(r) = i \, \zeta(r) \sinh {\kappa r\over 2} ~~~.
\label{sAxial1}
\eneq
The non-Abelian field strengths are
\be
F_{\mu\nu}^a &=&
 {1\over r^2}
  \ep_{\mu\nu b}  \hat x^a \hat x^b (\zeta(r)^2 -1)
+ {\kappa\over 2r} (\ep_{a\mu\nu} -  \ep_{\mu\nu b}  \hat x^a \hat x^b )    
  \zeta(r) \sinh{\kappa r\over 2} \cr
\noalign{\kern 10pt}
&& \hskip 4cm +{i\kappa \over 2 r}  (\delta^{a\nu} \hat
x^\mu-\delta^{a\mu}
\hat x^\nu)\zeta(r) \cosh {\kappa r\over 2}
\label{non-abelian field strength 2} ~.
\ee
    
The three equations reduce to
\be
&&{1\over r^2} {d\over dr} \Big( r^2 {d\over dr} h \Big) 
  - \lambda (h^2 - v^2) h -
{2\over r^2} \zeta(r)^2 h = 0 \cr
&&\zeta'' - {1\over r^2} (\zeta^2 -1) \zeta 
   - \Big( h^2 + {\kappa^2\over 4} \Big) \zeta = 0  ~~~.
\label{sAxial2}
\ee
The boundary conditions are (a) $h=0$ and $\zeta=1$ at $r=0$ and 
(b) $h=v$ and $e^{\kappa r/2} \zeta =0$ at $r=\infty$.
Eq.\ (\ref{sAxial2}) follows by minimizing  the integral
\beeq
S = \int_0^\infty dr \, \Bigg\{ (\zeta')^2 + {\kappa^2\over 4} \, \zeta^2
+ {1\over 2r^2} \,(\zeta^2 -1)^2 + \zeta^2 h^2
+ {r^2\over 2} (h')^2 + {\lambda r^2\over 4} (h^2-v^2)^2 \Bigg\} ~~.
\label{action4}
\eneq

The solution can be easily found numerically, as it follows from the
minimization of (\ref{action4}).  It has a finite Euclidean action
(\ref{action2}) and is depicted in Fig.\ 3.
 
\begin{figure}[ht]\centering
% \leavevmode 
\mbox{
\epsfysize=3.2in \epsfbox{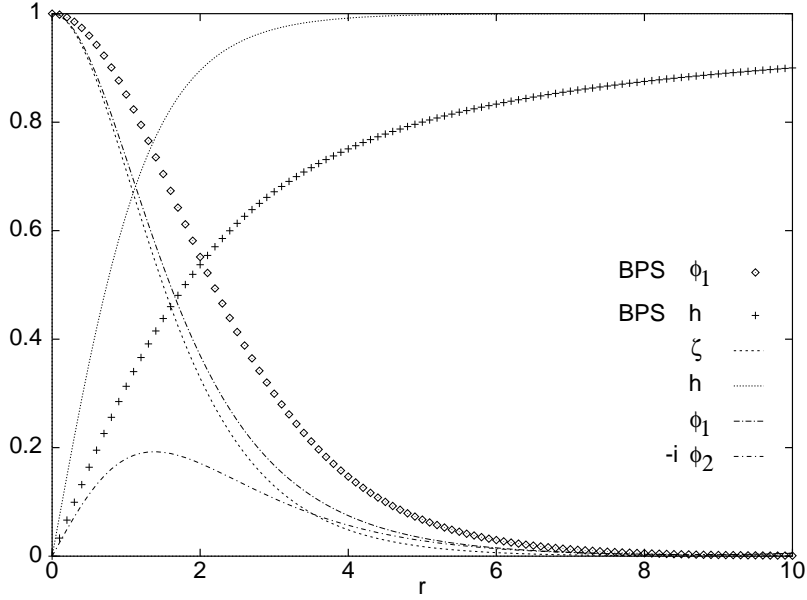}}
\caption{The solution in the radial gauge for $v=1$,  $\kappa =0.5$
and $\lambda=0.5$. The
solutions at arbitrary values of $\kappa$ and $\lambda$ are 
numerically generated using the exact known BPS solutions for 
$\phi_1$, BPS and $h_{BPS}$. In this plot we show the solutions for
$\zeta(r)$ and $h(r)$, and also $\phi_1(r)$ and $-i\phi_2(r)$.} 
\end{figure}

\bigskip

\noindent {\bf (ii) Unitary gauge ($\phi_2=0$)}

As remarked in the previous section there is no monopole solution
with a finite action in this gauge.  It may be of interest to
investigate a configuration which is related to the complex monopole 
solution in the radial gauge by a ``complex'' gauge transformation.
As $\phi_2/\phi_1 = i \tanh (\kappa r/2)$ is pure imaginary 
in the radial gauge, the gauge function
in (\ref{transform1}), which brings the solution in the unitary gauge,
is  $f= i \kappa r/2$.  Hence $A^{\rm unitary} = - i\kappa/2$, 
which gives a divergent action.  Of course the configuration thus 
obtained does not satisfy the equations in the unitary gauge.

\bigskip

\noindent {\bf (iii) Radiation gauge ($\dd_\mu A^a_\mu =0$)}

Again we look for a solution in which $\phi_2$ is pure imaginary:
$\phi_2 = i \phibar_2$ and $A = i \Abar $.   Equations to be solved
are
\be
&&{1\over r^2} {d\over dr} r^2 {d\over dr} h
-\lambda (h^2-v^2) h - {2\over r^2} (\phi_1^2 - \phibar_2^2)h =0 \cr
\noalign{\kern 5pt}
&&{d\over dr} \pmatrix{D\phi_1\cr D\phibar_2\cr} 
- \Big(\Abar + {\kappa\over 2} \Big) \pmatrix{D\phibar_2\cr D\phi_1\cr}
- \bigg\{  {1\over r^2} (\phi_1^2 - \phibar_2^2 -1) + h^2 + 
  {\kappa^2\over 4} \bigg\}  \pmatrix{\phi_1\cr \phibar_2\cr}  \cr
\noalign{\kern 10pt}
&&\hskip 4cm - \pmatrix{0\cr 1\cr} \int_r^\infty du \, {2\over u^2}
  \Big\{ \phibar_2 D\phi_1 - \phi_1 D\phibar_2 - {\kappa\over 2}
\Big\} =0
\label{EqMotion11}
\ee
where
\beeq
\pmatrix{D\phi_1\cr D\phibar_2\cr} =
 \pmatrix{\phi_1'\cr \phibar_2'\cr} 
- \Big( \Abar + {\kappa\over 2} \Big) \pmatrix{\phibar_2\cr \phi_1\cr} 
\next \Abar = {2\over r^2} \int_0^r du \, \phibar_2(u) ~~~.
\label{derivative1}
\eneq
It is not easy to find a simple ansatz for $\phi_1$ and $\phibar_2$
which solves the equations.  

Note that the equations in (\ref{EqMotion11}) are in a gauge covariant
form except for the last term in the second equation.  If one makes
a complex gauge transformation, promoting $f$ in (\ref{transform1}) to
a complex $f=i\bar f$, 
\beeq
\pmatrix{\phi_1\cr \phibar_2\cr} \go 
\pmatrix{ + \cosh \fbar & -\sinh \fbar\cr
        -\sinh \fbar & + \cosh\fbar\cr} \pmatrix{\phi_1\cr \phibar_2\cr} 
\next \Abar \go \Abar - \fbar' ~~~,
\label{transform4}
\eneq
then the equations remain the same except that the matrix factor
$(0,1)$ in the second equation in (\ref{EqMotion11}) is replaced 
by $(\sinh \fbar , \cosh\fbar)$.
Even if one chooses $\fbar$ such that the new $\Abar=0$, the 
last term remains.  The transformed equations are different from
the equations in the radial gauge.

The solution for Eqs.\ (\ref{EqMotion11}) and (\ref{derivative1}) need 
to be found numerically, which is left for future work.  Let us
suppose that there is a solution.  One has to ask if the solution 
is unique.  We explore ``complex'' Gribov copies of the solution.
As $\phi_2$ is pure imaginary, solutions to (\ref{radiation2}) 
necessarily become complex.  There is no ``real'' Gribov copy.
With $f=f_R + i f_I$ Eq.\ (\ref{radiation2}) becomes
\be
&&f_R'' + {2\over r} f_R' 
-{2\over r^2} \Big\{ \phi_1 \sin f_R \cosh f_I
- \phibar_2 \sin f_R \sinh f_I \Big\} = 0 \cr
\noalign{\kern 10pt}
&&f_I'' + {2\over r} f_I' 
-{2\over r^2} \Big\{ \phi_1 \cos f_R \sinh f_I
+ \phibar_2 (1 - \cos f_R \cosh f_I) \Big\} = 0 ~.
\label{radiation4}
\ee
Boundary conditions are $f_R(0) = f_I(0)=0$ and 
$f_R(\infty), f_I(\infty)=$~finite.  Although the meaning of these
solutions is not clear for $f_I(r)\not= 0$, we point out that solutions
satisfying $f_I(\infty)=0$ might have special role in the path
integral.  In view of (\ref{transform2}), such copies carry
the additional oscillatory factor $(4\pi i \kappa/g^2) f_R(\infty)$
in the path integral.  Examination of Eq.\ (\ref{radiation4}) with
representative $\phi_1$ and $\phibar_2$ shows that such a solution
is uniquely found with given $f_R'(0)$.  $f_R(\infty)$ is determined 
as a function of $f_R'(0)$.  The range of $f_R(\infty)$ is not
restricted to $[-\pi, \pi]$.  No cancellation is expected
in the path integral from these copies.

\sxn{Remarks on other works}

In this section, we wish to point out some differences between our work and
 ones found in the earlier literature, especially as  there appears to 
be some  confusion about the 
relevance of monopole configurations in the path integral.

\noindent (i) Pisarski's work

In ref.\ \cite{Pisarski}, the unitary gauge $\phi_2=0$ was adopted.
As remarked before, the boundary conditions cannot be satisfied
in this gauge for $\kappa\not= 0$.  In other words, solutions
obtained in this gauge after relaxing the boundary conditions
necessarily have an infinite action for $\kappa\not= 0$.  This is what
Pisarski found.  If one considers a monopole-antimonopole pair,
the action can be made finite.  The action is proportional to the
distance between the pair, which leads to the confinement picture
of monopole-antimonopole pairs.

We also remark again that Pisarski's solution in the unitary gauge is 
different from the configuration obtained from the solution in the
radial  gauge in Section 3-(i) by a gauge transformation.  

\noindent (ii) Affleck, Harvey, Palla, and Semenoff's work

It has been recognized in ref.\ \cite{Affleck} that gauge copies of
monopole solutions can lead to cancellation in the path integral.
As expressed in (\ref{transform2}), gauge copies yield an extra
factor $\exp \big\{ (4\pi i\kappa/g^2) f(\infty) \big\} 
= \exp \big\{ i n f(\infty) \big\}$ where the quantization condition
for $\kappa$ has been employed.   In ref.\ \cite{Affleck} the factor
$f(\infty)$ was promoted to a collective coordinate $\Lambda$.  It was argued
then that the integration over the collective coordinate $\Lambda$
from 0 to $2\pi$ gives a vanishing contribution when the monopole
number is non-vanishing.

However, this argument is incomplete.  As explained in Section 2-(iii), the possible range for $f(\infty)$ in the radiation gauge for
the monopole in the BPS limit is $[-3.98, 3.98]$.   No cancellation is 
expected.

\noindent (iii) Fradkin and Schaposnik's work

In ref. \cite{Fradkin}, the authors start with a gauge invariant theory with 
massive fermions in the Abelian theory.  The integration of fermion
degrees induces a Chern-Simons term, which leads to the deconfinement
of charges.  The authors have argued that there appears a linear
confining potential between  monopoles and antimonopoles so that 
a monopole gas becomes a dipole gas exhibiting no Debye screening
and destroying the confinement picture.
Further, they argue that the lack of the gauge invariance of the
Chern-Simons term makes gauge copies of monopole configurations leads
to the cancellation in the path integral.

Their argument is unsatisfactory in the  context of our work in
two respects.  First they consider monopole configurations
obtained in the absence of the Chern-Simons term and insert them into 
the Chern-Simons term to find implications.  The fact that the 
configurations do not solve the equations in the presence of the
Chern-Simons term leads to a linear potential among
monopoles and antimonopoles.  This argument is not consistent; one
should examine solutions in the full theory including the Chern-Simons 
term.

Secondly their argument is given for the
Abelian theory with monopole configurations carrying singularity
at their cores.  Although the authors claim that their
argument goes through for the non-Abelian case as well, there are
important differences between the two cases.  In the Abelian theory
with monopole background fields, the Chern-Simons term, after a
gauge transformation $\omega(\vx)$, yields an extra factor
$\exp \big\{ i \sum_a n_a \omega(\vx_a)/2 \big\}$ where $n_a$ and
$\vx_a$ are magnetic charge and position of the $a$-th monopole.
Hence Fradkin and Schaposnik argue that the integration of
$\omega(\vx)$ eliminates contributions of monopoles completely.
The factor $\omega(\vx_a)$ results from the singularity of the
Abelian monopole configuration at $\vx=\vx_a$.  In the non-Abelian
theory, however, monopole configurations are regular everywhere and
the Chern-Simons term produces a factor $\exp \big\{ inf(\infty)
\big\}$  (see (\ref{transform2})).  Only the value of the gauge potential
$f(r)$ at $r=\infty$ is important.  We have observed that $f(\infty)$
takes values in a limited range and no cancellation is expected.

\noindent (iv)  Diamantini, Sodano and Trugenberger's work

Diamantini et al.\ have  formulated the compact Maxwell-Chern-Simons
theory 
 in the dual theory on the lattice, in which the dual field
variable $f^\mu = \ep^{\mu\nu\rho} F_{\nu\rho}/2$ becomes   
fundamental.\cite{Diamantini}  In this compact $U(1)$ theory monopoles
naturally arise on the lattice. Diamatini et al.\ have found a complex
solution for a monopole-antimonopole pair.  The dual field $f^\mu$ has a
string singularity between the two poles. In the presence of the 
Chern-Simons term the string, carrying a magnetic flux, becomes
electrically charged.  As Henneaux and Teitelboim pointed
out,\cite{Henneaux} such a string becomes observable and has
a finite energy density so that the monopole-antimonopole pair is
linearly confined.

Although the confinement picture of monopoles in ref.\ \cite{Diamantini}
is consistent with refs.\ \cite{Pisarski, Affleck, Fradkin}, due caution
is necessary in extending the picture to non-Abelian theory.  The dual
theory in \cite{Diamantini} is entirely  Abelian, consisting of only
gauge fields.  In the Georgi-Glashow-Chern-Simons model, the solution is
regular everywhere. There is no place where an observable string
singularity enters.  It is not clear if the role of the Higgs
field in the continuum non-Abelian theory can be completely mimiced by
the  lattice structure in the dual theory.

\noindent (v) Jackiw and Pi's work

Jackiw and Pi \cite{Jackiw} have argued that the addition of the Chern-Simons
interaction destroys the topological excitations such as monopoles.
They parameterize $\phi_1= \rho \cos\theta$ and 
$\phi_2=\rho \sin\theta$.  The boundary conditions (\ref{BC1})
are $\rho(0)=1, \theta(0)=1$ and $\rho(\infty)=0$.  Under a gauge
transformation (\ref{transform1}), $\theta \go \theta - f$ and
$A \go A-f'$.  $\theta' - A$ is gauge invariant.

The authors  employ one of the classical equations of motion, 
Eq.\ (\ref{EqMotion2}),
to reduce the action, which is subsequently minimized.  However,
as remarked near the end of Section 2, Eq.\ (\ref{EqMotion2}) is 
incompatible with the boundary condition ensuring the finiteness of
the action.  Consequently all monopole configurations have an infinite
action in their formalism.

Indeed, Eq.\ (\ref{EqMotion2}) reads
$\rho^2 (\theta'- A - {1\over 2} i\kappa)= - {1\over 2} i\kappa$.
Upon utilizing this equation, the effective action (\ref{action2}) is
reduced to
\be
S_{JP} &=& {4\pi\over g^2} \int_0^\infty dr \, 
\Bigg\{  \rho'^2  + {\kappa^2\over 4} \Big( \rho^2 + {1\over \rho^2}
\Big)   \cr
&&\hskip 2cm + {1\over 2r^2} (\rho^2 - 1)^2 
+{r^2\over 2} h'^2 + h^2 \rho^2 + {\lambda r^2\over 4}
(h^2-v^2)^2 \Bigg\} 
\label{JPaction}
\ee
The presence of the $1/\rho^2$, $h^2\rho^2$, and $r^2(h^2-v^2)^2$
 terms makes it
impossible to have a configuration of a finite action.

In quantum theory  a gauge is fixed.
In the radial gauge $A=0$, for instance,  one of the classical equation, 
 Eq.\ (\ref{EqMotion2}), which is derived by varying $A$,  does not
follow. Hence the relation $\rho^2 (\theta'- A - {1\over 2} i\kappa)= -
{1\over 2} i\kappa$ should not be used to simplify the action.

\noindent (vi)  More subtleties

There remains subtle delicacy in defining the quantum theory of 
the Chern-Simons theory.   We have started with the Faddeev-Popov 
formula (\ref{PIformula1}).  If one picks a radial gauge 
$F(A)= x^\mu A^a_\mu=0$, then there is a complex monopole solution
which extremizes the restricted action, namely the action in the given
gauge slice.  What happens, say, in the temporal gauge?  The answer is
not clear.  As explained in Section 4, solutions look different,
depending on the gauge chosen. Two operations in the path integral,
fixing a gauge and finding configurations which extremize the action, do
not commute with each other in the Chern-Simons theory.
In the above papers by Pisarski and by Jackiw and Pi the action is
extremized with respect to arbitrary variations of gauge fields, and
then a gauge is picked.  This procedure yields one more equation to
be solved, and in general this equation turns out   incompatible 
with the finiteness of the action for monopoles.

In QED or QCD the order of the two operations does not matter.
For instance, instanton solutions in QCD can be found in any gauge.
The apparent noncommutability of the two operations in the Chern-Simons
theory is traced back to the gauge non-invariance of the Chern-Simons
term particularly in the monopole background.  In the original
derivation of the Faddeev-Popov formula (\ref{PIformula1})  the 
gauge invariance of the action was assumed.  The formula
(\ref{PIformula1}) needs to be scrutinized in the Chern-Simons theory.
There is also ambiguity in the definition of the Chern-Simons
term as pointed out by Deser et al.\cite{Deser2}  Further investigation
is necessary.

\sxn{Chern-Simons-Higgs theory}

In the absence of Yang-Mills term one obtains
the Chern-Simons-Higgs theory. Although the pure Chern-Simons theory
defines a topological field theory, 
one obtains a dynamical theory after matter couplings are introduced.
The equations of motion for the gauge fields are
first order in derivatives which makes the theory easier to handle
at least at the classical level.
The issue whether this theory makes sense or not at the quantum level was
addressed by Tan {\it et. al.} in ref. \cite{Tan}.  Using the two loop
effective potential in the dimensional regularization scheme, it has been
shown that  one has to start with the
Yang-Mills term, but the limit of vanishing Yang-Mills term exists after
renormalization.  In this section we are mainly interested in the 
Chern-Simons-Higgs theory at the tree level.
 
Monopoles in CS-Higgs theory was  discussed in references \cite{Lee} and
 \cite{Edelstein}. 
Lee showed that  instantons of an infinite action 
induce an effective
vertex which break the global part of the $ U(1)$ gauge symmetry and lead
to non-conservation of charge. 
Edelstein and  Schaposnik showed that there are no monopole  solutions of a 
finite action. Using the vacuum equations of motion they showed that magnetic
field is everywhere orthogonal to the Higgs field in the isospin space in
contrast to the 't Hooft-Polyakov hedgehog solution where two fields 
are parallel. This 
orthogonality of the magnetic field and  Higgs field forces the
$U(1)$ field strength to vanish for finite action monopoles.
In this section we are going to reproduce the same conclusions as
in \cite{Edelstein} in a way that parallels our discussions in the
previous parts of this paper. 

The action in terms of  the monopole ansatz is
\be
S &=& {4\pi\over g^2} \int_0^\infty dr \, \Bigg\{
+i \kappa \Big[ \phi_1' \phi_2 - \phi_2'(\phi_1 -1) + A(\phi_1^2 +
\phi_2^2 -1) \Big] \cr
&&\hskip 2cm 
+{r^2\over 2} h'^2 + h^2 (\phi_1^2+\phi_2^2) + {\lambda r^2\over 4}
(h^2-v^2)^2 \Bigg\} 
\label{action5}
\ee
The finiteness of the action demands boundary conditions
$h=v, \phi_1=\phi_2=A=0$ at $r=\infty$, but imposes no condition
at $r=0$.  The regularity of the configuration at the origin 
requires $h=\phi_2=0, \phi_1=1$ at $r=0$.

If all $A, \phi_1, \phi_2$ and $h$ are regarded as independent
variables, then equations of motion are
\be
&&\phi_1^2 + \phi_2^2 - 1 =0 \label{CSeq1} \\
&&i\kappa (\phi_2' - A\phi_1) - h^2 \phi_1 =0 \label{CSeq2} \\
&&i\kappa (\phi_1' + A\phi_1) + h^2 \phi_2 =0 \label{CSeq3} \\
&&(r^2h')' -2(\phi_1^2 +\phi_2^2) h -\lambda(h^2-v^2)r^2 h =0 ~~.
  \label{CSeq4}
\ee
Clearly the first equation (\ref{CSeq1}) is incompatible with the 
boundary conditions at $r=0$ and $\infty$.

However, a gauge is fixed in quantum theory.  Take the radial gauge
$A=0$.  Then equations of motion are (\ref{CSeq2}),  (\ref{CSeq3}),
and  (\ref{CSeq4}) where $A$ is set to be 0.  In the Jackiw-Pi
parameterization $\phi_1 + i\phi_2 =\rho e^{i\theta}$, Eqs.\ 
(\ref{CSeq2}) and  (\ref{CSeq3}) become
\beeq
(i\kappa \theta' + h^2) \rho = 0 \next i\kappa \rho' =0 ~~~.
\label{CSeq5}
\eneq
The second equation implies that $\rho$ is constant, which conflicts
with the boundary conditions.  Hence there is no regular monopole
configuration of a finite action. In fact equations of motion totally
eliminate the spherically symmetric gauge fields and the resulting action 
is that of a Higgs field alone with the action given by:
\be
S &=& {4\pi\over g^2} \int_0^\infty dr \, \Bigg\{
{r^2\over 2} h'^2   + {\lambda r^2\over 4}
(h^2-v^2)^2 \Bigg\}
\label{action6}
\ee

\sxn{Conclusions and discussions}

We have seen that in the Georgi-Glashow-Chern-Simons theory, complex
monopoles exist, and that they have a non-vanishing contribution to the 
path integral. As we have shown,  the cleanest way to see this is in the
radial gauge. The action is  minimized by complex solutions, and is real
and finite. Furthermore, the solutions 
 have the usual  characteristics of monopoles.  They 
have $U(1)$ field strength 
given by $F_{\mu\nu}= -\epsilon_{\mu\nu}\hat x^a / r^2$ and  
mass $\sim 1/g^2$.  As a consequence, the 
long range order in the Higgs vacuum is destroyed. However, 
we must recognize that we are far from understanding this theory
at quantum level, or beyond semiclassical approximation. 
The understanding of the quantum theory is obscured by the 
gauge non-invariance of the Chern-Simons term.

We started with a theory with compact U(1) symmetry, where by definition  
the gauge transformation parameter (which we called $f$) is originally 
a real-valued
function. However, in discussing the Gribov copy problem in the
radiation gauge, we have looked for complex field configurations
which are related to each other by complex $f$.

Curiously enough, the gauge invariant 
part of the action (\ie everything other than the 
CS term) is still invariant under the transformation with complex
$f$.   At the tree level there are many saddle points in the complex field
configuration  space which are related by these complex $f$'s.
At the one loop level, however, the effective action would not be
invariant under complex $f$ transformations. The physical interpretation
of $f$ being complex is not  obvious at all.

If we had restricted
the gauge parameter $f$ in this theory to be
real and not allow complex $f$, then the absence of real solutions to
the  Gribov  equation in the radiation gauge would lead to 
no Gribov copies. 

We have adopted the view that  complex solutions to Gribov's equation 
correspond to generalized Gribov copies of  complex saddle points. 
We understand that this is a question not completely settled, and
warrants further investigation. Meanwhile we conclude that within
our semi-classical approximation, it appears that  summation over 
the Gribov copies
(or integrating over the collective coordinates) of the complex 
monopole solutions does not lead to the cancellation of the monopole 
contribution.  What if quantum
corrections  to the Jacobian of the Gribov copies somehow cancel the
effect after all?   This is one of the questions which we can not answer
in this model unless we learn how to  go beyond the semi-classical
approximation. We remark that
one could raise the same objection for the real monopole case where 
it has been argued in the literature that 
the integral cancels.   Recall that the issues of the non-invariance 
of the CS
term,  and the problems associated with the quantization are
irrespective of whether the  monopole is real or complex.

In closing, let us remark that complex gauge field configurations have been studied
before in the literature \cite{Wu}-\cite{Asorey}.  In particular Wu and
Yang  have given a
prescription of how complex gauge fields in $SU(2)$ theory can be
converted to real gauge fields for the group $SL(2,C)$. Witten
 describes a way to quantize  theories with non-compact
gauge groups. He shows that Chern-Simons theory with the group $SL(2,C)$
is equivalent to the 2+1 dimensional quantum gravity. According to
the analysis of ref.\ \cite{Witten}, one can obtain  
the Einstein-Hilbert
theory  both in the de-Sitter and  anti-de-Sitter spaces. Further research
is needed to understand  how one can generalize this description to the
space where cosmological constant is zero  and how one can couple matter
to gravity in this language. Making contact between these
three-dimensional gravity related considerations and the present work is
the subject of a separate study to be presented elsewhere.
 
\vskip 1cm

\leftline{\bf Acknowledgements}

The authors would like to thank Bob Pisarski and 
Pierre van Baal for useful discussions.  They are also grateful
to Roman Jackiw and So-Young Pi for many enlightening remarks
and for pointing out an error in the original version of the paper.  
This work was supported in part 
by  the U.S.\ Department of Energy under contracts DE-FG02-94ER-40823.
One of the authors (Y.H.) would like to thank the Yukawa Institute
for Theoretical Physics for its hospitality where the final part 
of the work was done in summer 1998.

\vskip 1cm

\leftline{\bf References}  

\renewenvironment{thebibliography}[1]
        {\begin{list}{[$\,$\arabic{enumi}$\,$]}  % {\arabic{enumi}.}
        {\usecounter{enumi}\setlength{\parsep}{0pt}
         \setlength{\itemsep}{0pt}  \renewcommand{\baselinestretch}{1.2}
         \settowidth
        {\labelwidth}{#1 ~ ~}\sloppy}}{\end{list}}

\myend
\begin{thebibliography}{99}

\small

\bibitem{Polyakov}
A.\ M.\ Polyakov, \Journal{\PLB}{59}{1975}{80};
\Journal{\NPB}{120}{1977}{429}. 

\bibitem{Deser}
S.\ Deser, R.\ Jackiw and S.\ Templeton, \Journal{\PRL}{48}{1982}{976}.

\bibitem{Pisarski}
R.\ D.\ Pisarski, \Journal{\PRD}{34}{1986}{3851}. 

\bibitem{D'Hoker} 
E.\ D.\ D'Hoker and L.\ Vinet, \Journal{\AP}{162}{1985}{413}. 

\bibitem{Affleck}
I.\ Affleck, J.\ Harvey , L.\ Palla and G.\ Semenoff, 
\Journal{\NPB}{328}{1989}{575}.

\bibitem{Fradkin}
E.\ Fradkin and F. A.\ Schaposnik, \Journal{\PRL}{66}{1991}{276}.

\bibitem{Diamantini}
M.\ C.\ Diamantini, P.\ Sodano and C.\ A.\ Trugenberger,
\Journal{\PRL}{71}{1993}{1969}.

\bibitem{Hosotani}
Y.\ Hosotani, \Journal{\PLB}{69}{1977}{499}.

\bibitem{Diamantini2}
M.\ C.\ Diamantini, P.\ Sodano and C.\ A.\ Trugenberger,
\Journal{\NPB}{474}{1996}{641}.


\bibitem{tHooft} G.\ 't Hooft, \Journal{\NPB}{79}{1974}{276}.


\bibitem{Gribov} V.\ N.\ Gribov, \Journal{\NPB}{139}{1978}{1}.

\bibitem{Dunne}
G. V.\ Dunne, R.\ Jackiw and C. A.\ Trugenberger,
\Journal{\AP}{194}{1989}{197}.

\bibitem{Henneaux}
M.\ Henneaux and C.\ Teitelboim,  \Journal{\PRL}{56}{1986}{689}.

\bibitem{Jackiw} 
R.\ Jackiw and S.Y.\  Pi, \Journal{\PLB}{423}{1998}{364}

\bibitem{Tan} 
P-N.\ Tan, B.\ Tekin and Y.\ Hosotani, \Journal{\PLB}{388}{1996}{611};
\Journal{\NPB}{502}{1997}{483}.

\bibitem{Lee}
K.\ Lee, \Journal{\NPB}{373}{1992}{735}.

\bibitem{Edelstein} J.\ D.\ Edelstein and F.\ A.\ Schaposnik, 
\Journal{\NPB}{425}{1994}{137}.

\bibitem{Deser2}
S.\ Deser, L.\ Griguolo and D.\ Seminara, preprint hep-th/9712132.

\bibitem{Wu}
T\ T.\ Wu and C.N.\ Yang, \Journal{\PRD}{13}{1976}{3233}.


\bibitem{Witten} E.\ Witten, \Journal{\NPB}{311}{1988}{46};
\Journal{\cmp}{137}{1991}{29}.


\bibitem{Oh}
C.\ H.\ Oh , L.C.\ Sia and E. C. G.\ Sudarshan, 
\Journal{\PLB}{248}{1990}{295}.

\bibitem{Asorey}
M.\ Asorey,  F.\ Falceto, J.\ L.\ Lopez and 
G.\ Luzon, \Journal{\PLB}{349}{1995}{125}.



\end{thebibliography}
